\documentclass{egpubl}
\usepackage{pg2023}

\SpecialIssuePaper         

\CGFStandardLicense

\usepackage[T1]{fontenc}
\usepackage{dfadobe}  

\usepackage{cite}  
\BibtexOrBiblatex
\electronicVersion
\PrintedOrElectronic
\ifpdf \usepackage[pdftex]{graphicx} \pdfcompresslevel=9
\else \usepackage[dvips]{graphicx} \fi

\usepackage{egweblnk} 

\usepackage{amsfonts,amssymb}
\usepackage{multirow}
\usepackage{diagbox}
\usepackage{booktabs}
\usepackage{amsmath}
\usepackage[colorlinks,linkcolor=blue]{hyperref}

\title[GA-Sketching: Shape Modeling from Multi-View Sketching with Geometry-Aligned Deep Implicit Functions]%
      {GA-Sketching: Shape Modeling from Multi-View Sketching with Geometry-Aligned Deep Implicit Functions}

\author[J. Zhou et al.]
{\parbox{\textwidth}{\centering 
        Jie Zhou$^{1}$\orcid{0000-0002-3836-4163},
        Zhongjin Luo$^{2}$,
        Qian Yu$^{3}$,
        Xiaoguang Han$^{2}$,
        and Hongbo Fu\thanks{Corresponding author: Hongbo Fu (E-mail: hongbofu@cityu.edu.hk).}$^{1}$\orcid{0000-0002-0284-726X}
        }
        \\
 {\parbox{\textwidth}{\centering $^1$ City University of Hong Kong\\
         $^2$ The Chinese University of Hong Kong, Shenzhen \\
         $^3$ Beihang University
       }
 }
}

\begin{document}

\teaser{
 \includegraphics[width=0.99\linewidth]{images/fig_teaser.png}
 \centering
  \caption{\small{We present GA-Sketching, a novel interactive system designed for novice users to create 3D objects through an iterative multi-view sketching process. Users can intuitively customize desired 3D objects by progressively drawing sketches from any number of viewpoints.}}
\label{fig:teaser}
}

\maketitle
\begin{abstract}
  Sketch-based shape modeling aims to bridge the gap between 2D drawing and 3D modeling by providing an intuitive and accessible approach to create 3D shapes from 2D sketches. However, existing methods still suffer from limitations in reconstruction quality and multi-view interaction friendliness, hindering their practical application. This paper proposes a faithful and user-friendly iterative solution to tackle these limitations by learning geometry-aligned deep implicit functions from one or multiple sketches. Our method lifts 2D sketches to volume-based feature tensors, which align strongly with the output 3D shape, enabling accurate reconstruction and faithful editing. Such a geometry-aligned feature encoding technique is well-suited to iterative modeling since features from different viewpoints can be easily memorized or aggregated. Based on these advantages, we design a unified interactive system for sketch-based shape modeling. It enables users to generate the desired geometry iteratively by drawing sketches from any number of viewpoints. In addition, it allows users to edit the generated surface by making a few local modifications. We demonstrate the effectiveness and practicality of our method with extensive experiments and user studies, where we found that our method outperformed existing methods in terms of accuracy, efficiency, and user satisfaction. The source code of this project is available at \href{https://github.com/LordLiang/GA-Sketching}{https://github.com/LordLiang/GA-Sketching}.

\end{abstract}

\section{Introduction}
  Sketching has been an intuitive way for humans to depict 3D shapes since prehistoric times. In modern society, sketches are still widely used by various users, including fashion designers and architectural engineers, to express their ideas. Designers also repeatedly modify details and iterate between sketches and corresponding 3D models until the desired shapes are achieved. While shape modeling from a single sketch has been well explored in recent years~\cite{zhang2021sketch2model, guillard2021sketch2mesh, gao2022sketchsampler, zheng2023lasdiffusion}, a single sketch may not capture all geometric details due to occluded regions and the inherent ambiguity of sparse line drawings. Sometimes, it is difficult for humans to imagine 3D structures from a single sketch. To reduce inherent ambiguity and achieve more visual cues, seeking help from multi-view sketches is an easy solution that comes to mind. Shape modeling from multi-view sketches involves generating a 3D shape using sketches drawn from different viewpoints. Existing methods for shape modeling from multi-view sketches often utilize multi-branch inputs with fixed viewpoints~\cite{lun20173d, zhong2020towards, du2020sanihead} or employ an iterative refinement strategy~\cite{delanoy20183d, chowdhury2022garment} to improve modeling quality. However, these methods still suffer from limitations such as the inability to retain faithful details depicted by input sketches~\cite{delanoy20183d, chowdhury2022garment} and the inefficiency of multi-view interaction~\cite{lun20173d, li2018robust}. These methods may work well to reconstruct simple shape prototypes but are far from a practical design tool. 

  We believe that a successful solution for sketch-based modeling should satisfy two properties: {\emph{faithfulness}} and \emph{user-friendliness}. \emph{Faithfulness} refers to high-quality reconstruction and faithful local editing. The generated shapes should be as consistent as possible with the input sketches. \emph{User-friendliness} entails a simple and intuitive interface with readily available tools, enabling users to create and modify the desired 3D models promptly. Common editing operations, such as adding, removing, or changing a component, should generate accurate modification for the editing region and keep the remaining region unaffected. We aim to provide an interactive system for sketch-based shape modeling that meets the above two properties. 

  We speculate that the low faithfulness of previous methods is due to their adopted feature encoding technique. 
  Some existing methods~\cite{lun20173d, delanoy20183d, chowdhury2022garment} often encode the input sketch as a global feature, typically, a compact latent code vector, which is not strictly aligned with the 2D input, failing to preserve spatial details of the input sketches. Furthermore, due to the nature of dimension reduction, global feature encoding is a bottleneck for shape editing. It can be challenging for a flattened vector to provide local controllability and strong generalizability to unseen shape variations. To address this issue, some methods~\cite{saito2019pifu, xu2019disn} extract pixel-aligned features to preserve local details. LAS-Diffusion~\cite{zheng2023lasdiffusion}, a concurrent work with ours, employs local patch features to guide feature learning. Slightly differently, we encode an input sketch as geometry-aligned features by using a predicted depth map as a prior to preserve more accurate local details.
  The low-quality reconstruction of previous methods is also possibly due to their adopted shape representations.
  Recently, deep implicit functions have emerged as promising methods for representing complex and irregular surfaces~\cite{mescheder2019occupancy, park2019deepsdf, chibane20ifnet}. Compared with other shape representations (image-based~\cite{lun20173d, li2018robust, zhong2020towards}, mesh-based~\cite{du2020sanihead, zhang2021sketch2model}, voxel-based~\cite{delanoy20183d, delanoy2019combining, jin2020contour}, and point-based~\cite{gao2022sketchsampler, wang20233d}), deep implicit functions are lightweight and not limited in resolution, making them an attractive option for surface reconstruction.
  Based on the above discussion, we propose a novel method to generate 3D shapes from 2D sketches by learning geometry-aligned deep implicit functions. We encode the input sketches as volume-based feature tensors strongly aligned with both input 2D sketches and output 3D surfaces. Then, we learn deep implicit functions from these feature tensors to generate the final continuous surfaces. Compared with the global latent code vector, the volume-based feature tensor is naturally capable of memorizing and fusing features from different views, which is profitable for faithful local editing and iterative modeling.
  
  Dealing with multi-view inputs is non-trivial as it involves many design options, such as the input order and aggregation strategies. Some methods~\cite{lun20173d, zhong2020towards, du2020sanihead} directly take a complete set of multi-view sketches with fixed viewpoints as inputs, which means that sketches drawn from different views are fed into the model simultaneously. Such a setting is inflexible as users need to re-draw multi-view sketches once they change their ideas. Therefore, we adopt an iterative modeling pipeline by feeding multi-view sketches sequentially. When aggregating features from different viewpoints, some image-based methods~\cite{huang2018deepmvs, saito2019pifu} handle this issue by simple pooling operations, e.g., max/average pooling, which are unsuitable for iterative modeling, since the later sketches are more important and thus should have higher weights. Besides, these methods struggle to preserve view consistency among different sketches as the misalignment in each iteration will accumulate. Some incremental methods~\cite{delanoy20183d, chowdhury2022garment} use iterative refinement strategies to handle this issue, but still face challenges in achieving accurate and faithful results. To handle the above issues, we design an iterative feature aggregation module that aggregates two adjacent feature tensors into an integrated one through an iterative process. Additionally, we have implemented a masked editing strategy to ensure the consistency of the reconstruction results with the most recent sketch.
 
  Existing sketch-based modeling systems typically support shape editing with a series of \textit{predefined} operations, thus lacking flexibility. For example, SimpModeling~\cite{luo2021simpmodeling} provided several shape editing function buttons, such as extrusion, add-control-curve and handle-deformation. DeepSketch2Face~\cite{han2017deepsketch2face} designed 10 gesture-based interactions to allow users to manipulate initial face models generated from the first sketch. However, preset functions may not satisfy users' requirements for editing. Moreover, remembering these functions could be a burden for users. Therefore, we design a unified interactive system for 3D shape generation and editing to provide users with an intuitive and flexible interface, as shown in Figure~\ref{fig:teaser}. Our system not only allows users to generate a 3D shape from drawn sketches but also supports them in editing the shape in a free-form way. That is, users can edit by erasing unsatisfactory parts and directly drawing new ones. It is worth noting that our system can automatically render a reference sketch from the 3D shape for users to edit, relieving them of the burden of redrawing a sketch from scratch.
 
  To summarize, we make the following contributions:
  \begin{itemize}
  \item We propose a novel iterative solution to generate 3D shapes from 2D sketches by learning geometry-aligned deep implicit functions. To our knowledge, our method is the first to use geometry-aligned feature encoding for multi-view sketch-based shape modeling.
  \item  We design a unified interactive system for sketch-based shape generation and editing. {Our system enables users to generate or edit a 3D shape by drawing and modifying sketches from arbitrary viewpoints.}
  \end{itemize}


\section{Related Work}
  We will focus on discussing the existing efforts for multi-view sketch-based modeling. 
  
  \subsection{Traditional Sketch-based Modeling} 
  Since sketch-based modeling has a long history dating back to early Sutherland’s SketchPad system~\cite{sutherland1964sketch}, it is worthwhile to revisit some classic traditional methods. Starting with Teddy~\cite{igarashi1999teddy}, traditional methods often involved algorithms for recognizing and interpreting 2D sketches as closed polylines~\cite{karpenko2006smoothsketch} or implicit functions~\cite{karpenko2002free, tai2004prototype, schmidt2007shapeshop} and then using these interpretations to create smooth 3D models. To reduce the ambiguity of 2D sketches, FiberMesh~\cite{nealen2007fibermesh} allows users to model free-form surfaces by sketching and manipulating 3D curves. Schmidt et al.~\cite{schmidt2008sketch} proposed layered procedural surface editing operations to make 3D modeling more efficient. However, these traditional systems may produce over-smooth or unrealistic results and require significant manual effort to specify complex geometry, due to the lack of domain knowledge of 3D objects.
  
  \subsection{Data-driven Sketch-based Modeling}
  In recent years, data-driven methods have emerged, enabling the direct creation of 3D models from 2D sketches without geometric constraints or shape grammars. Early data-driven methods~\cite{xie2013sketch, fan2013modeling} relied on an input sketch as a query to retrieve the most similar pre-existing models from a shape repository and then deformed these models to better fit the contours specified in the sketch. However, these methods are easily limited by the size and diversity of the shape repository. Some methods~\cite{huang2016shape, han2017deepsketch2face} learned CNN-based deep regression networks to map sketches to parameters, which are used to deform a morphable model. Later methods moved beyond morphable models and utilized deep networks to learn various shape representations directly. For example, image-based representations~\cite{lun20173d, li2018robust, zhong2020towards} inferred the depth and normal maps representing the underlying surface, followed by complicated post-processing to fuse these maps to a completed 3D model. Mesh-based representations~\cite{du2020sanihead} typically learned to deform an initial convex template and hence failed to represent complex topologies. Volumetric representations~\cite{delanoy20183d, delanoy2019combining, jin2020contour} were computationally expensive and memory-intensive, severely limiting the resolution of the output shapes. Recent advances in the use of deep implicit functions~\cite{luo2021simpmodeling, chowdhury2022garment} have shown great promise in representing complex and irregular surfaces. We combine deep implicit functions with geometry-aligned feature encoding to represent implicit surfaces, which has not been used in previous multi-view sketch-based modeling.


\section{User Interface}
  In this section, we will introduce the user interface of our iterative multi-view sketch-based modeling system.
  \begin{figure}[!t]
  \centering
  \includegraphics[width=0.9\linewidth]{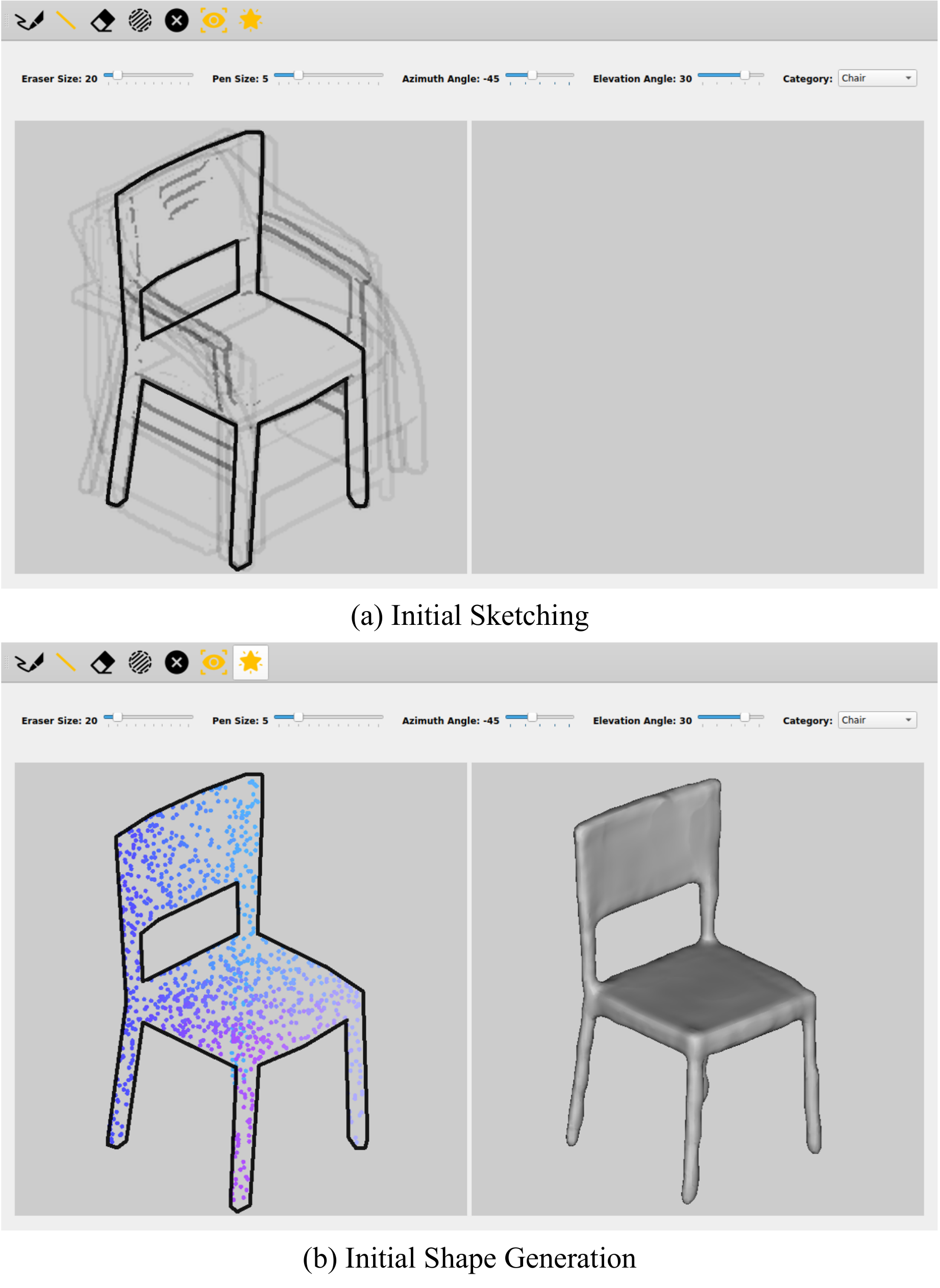}
  \caption{\small{The user interface of our system. (a) An initial sketch of a chair drawn by the user. (b) Click the $\bigstar$ button to generate the initial 3D shape. Some colorful points sampled from the initial shape are visualized to provide spatial perception for subsequent editing.}}
  \label{fig:ui}
\end{figure}
  \begin{figure*}[!t]
  \centering
  \includegraphics[width=0.9\linewidth]{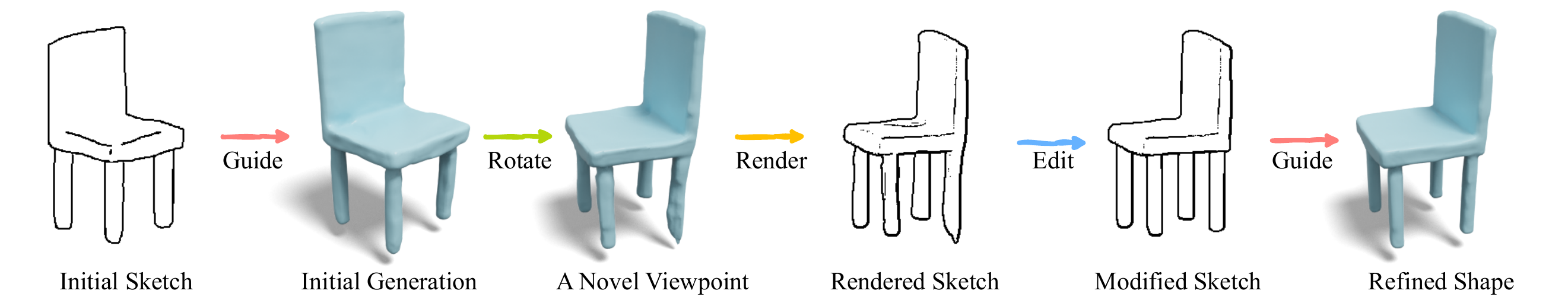}
  \caption{\small{The illustration of generation/refinement operations supported by our interactive system. The user can first draw an initial sketch and generate an initial 3D shape. Next the user can rotate the initial generation to a novel viewpoint and a reference sketch will be rendered automatically. The user can do some modification on the reference sketch to refine the shape.}}
  \label{fig:refine_workflow}
\end{figure*}

  \begin{figure*}[!t]
  \centering
  \includegraphics[width=0.8\linewidth]{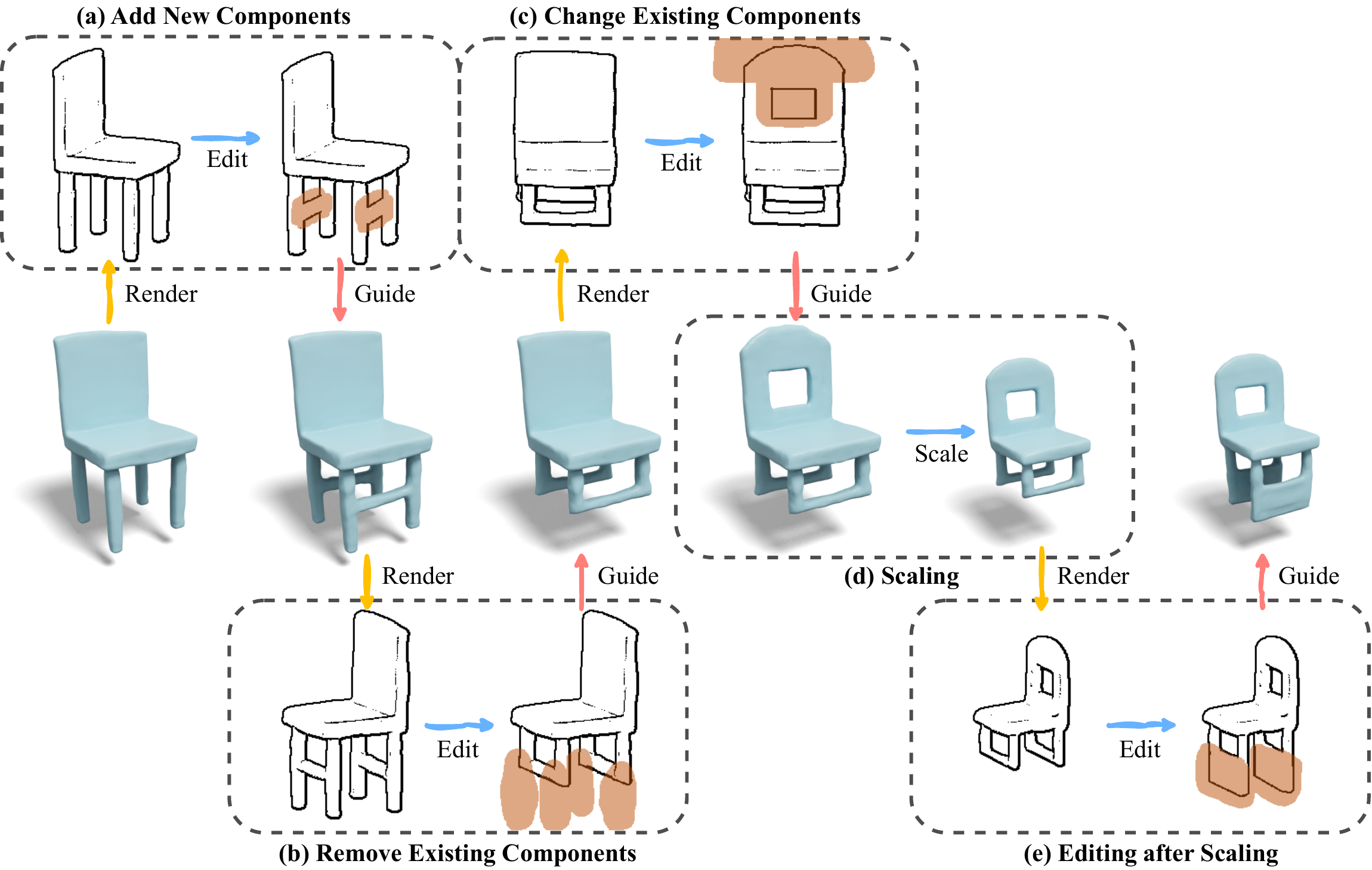}
  \caption{\small{The illustration of editing operations supported by our interactive system. Users need first mark out a local region for editing and then modify the geometry by sketching within this region. Here we show a representative iterative editing process: (a) Add New Components, (b) Remove Existing Components, (c) Change Existing Components, (d) Scaling and (e) Editing after Scaling.}}
  \label{fig:edit_workflow}
\end{figure*}

\subsection{User Interface Design}
  {Figure~\ref{fig:ui}} shows the main user interface with two working spaces. The left is a canvas space for sketch drawing, and the right is a viewer space to examine the reconstructed shape. Users can freely change viewpoints for these two working spaces. Considering users' drawing habits, we limit the elevation angle of the canvas space between $-15^{\circ}$ and $45^{\circ}$ (though our method supports any elevation angle) and allow the azimuth angle to cover all $360^{\circ}$. Inspired by ShadowDraw~\cite{lee2011shadowdraw}, we use a shadow sketch as a background to help users to draw the first sketch. The toolbar at the top provides four basic drawing tools (including free-form curve, straight line, eraser, and editing mask) and three function buttons, including the clear button for clearing the drawing, the lock button for locking and unlocking the viewpoint, and the generation button for 3D shape generation. Below the toolbar are four sliders to adjust the brush sizes and angles of shadow images, updated based on a user-selected shape category.
 
\subsection{Sketch-based Shape Generation}
  After completing the first sketch, the user can click the generation button to get the current reconstructed shape. Some colorful points sampled from the reconstructed surface is visualized to provide spatial perception. The user can rotate the reconstructed mesh freely in the viewer space to check whether some components need to be refined. This is often needed to address the ambiguity in sketches, particularly single-view sketches. After the user changes to a new viewpoint by rotating the current sketch, the system will generate a reference sketch from the current shape under this new viewpoint. The user can modify the reference sketch and click the generation button again to get a refined shape, as illustrated in Figure~\ref{fig:refine_workflow}. If the reference sketch is far from the desired sketch under the novel viewpoint, the user can click the clear button to empty the canvas and draw a sketch from scratch.
  
\subsection{Sketch-based Shape Editing} 
  It is important to distinguish between shape editing and shape refinement, as both involve making changes to the current shape. Shape refinement refers to the slight modification of the coarse generation from a novel viewpoint, with the goal of resolving any ambiguity during shape inference from sketches. On the other hand, shape editing involves changing the geometric structure of the current shape, which typically occurs when designers want to modify their original design ideas, e.g., removing a leg from a chair, adding a tail wing to an airplane, or changing the upper edge of a chair back from a straight line to a curved one.
  
  Our method supports the following shape editing operations, as shown in {Figure~\ref{fig:edit_workflow}} (a)-(c): adding new components, removing existing components, or changing existing components. Specifically, we ask the user to mark out a 2D binary mask and modify the sketch only within the editing region. In addition, our method also supports proportionally global scaling up/down, as shown in Figure~\ref{fig:edit_workflow} (d). The user can use the mouse wheel to zoom the current reference sketch in or out, and then our system will scale up/down the volume-based feature tensors correspondingly by trilinear interpolation. The scaled shape can be seamlessly integrated into the next refinement or editing step, as shown in Figure~\ref{fig:edit_workflow} (e).
  
\begin{figure*}[!t]
  \centering
  \includegraphics[width=0.95\linewidth]{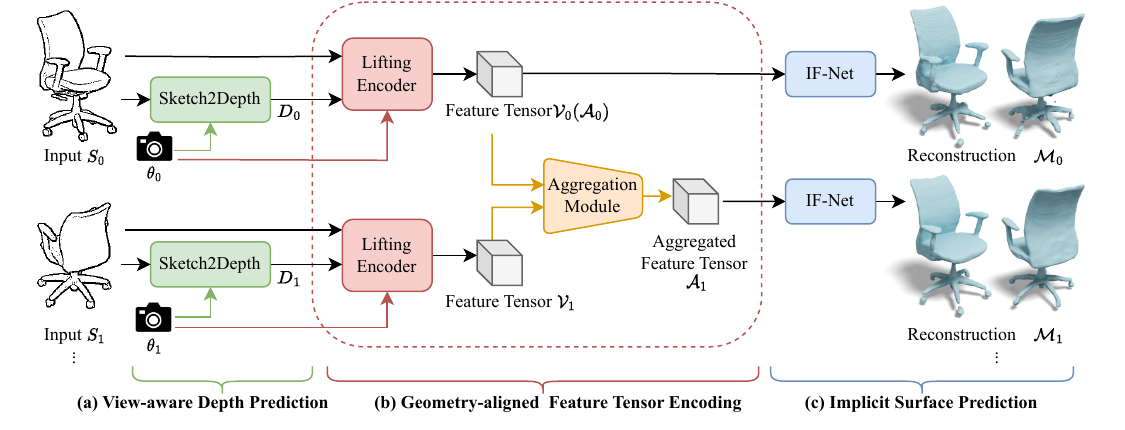}
  \caption{\small{
  Our method includes three stages: view-aware depth prediction, geometry-aligned feature tensor encoding, and implicit surface prediction. Firstly, the input sketches are fed into a Sketch2Depth translator to generate depth prediction for information richness. Then, we feed the predicted depths as well as the input sketches into a lifting encoder to generate geometry-aligned feature tensors. The feature tensors from different viewpoints are iteratively aggregated using our aggregation module. Finally, the aggregated feature tensor is fed into an IF-Net to predict occupancy values for query points.}}
  \label{fig:network}
\end{figure*}

\section{Method}
  Given one or more input sketches, our method aims to generate a high-quality 3D shape and allows fine-grained shape editing. The pipeline of our method mainly consists of three stages, as shown in Figure~\ref{fig:network}. Firstly, the input sketches are fed into a Sketch2Depth translator to generate depth prediction for information richness (Section~\ref{sec:prior}). Then we concatenate predicted depths with the input sketches and feed them into a lifting encoder to generate geometry-aligned feature tensors. These feature tensors are then iteratively aggregated using an aggregation module (Section~\ref{sec:encoding}). Subsequently, the aggregated feature tensor is fed into an IF-Net~\cite{chibane20ifnet} to predict an occupancy value for each query point (Section~\ref{sec:prediction}). In this section, we will introduce these three stages in detail.
  
\subsection{View-aware Depth Prediction}
\label{sec:prior}
  Due to a large domain gap between 2D sketches and 3D shapes, we attempt to compensate the missing information by translating an input sketch $S_i$ to a predicted depth $D_i$. Our Sketch2Depth translator is based on a classic U-Net~\cite{ronneberger2015u} structure. We add the camera parameter $\theta_i$ to the bottom-most of the network to add view awareness. In order to take full advantage of the characteristics of iterative modeling, we also enhance the Sketch2Depth translator by introducing the rendered depth as a reference. This is a hidden trick used in our interactive system to enhance the stability of depth prediction. Let $\mathcal{M}_{i-1}$ be the intermediate shape reconstruction and $S_i$ is the current sketch from the $i$-th view. We use a depth renderer $\mathcal{R}_D$~\cite{johnson2020accelerating} to render the depth $d_i$ from $\mathcal{M}_{i-1}$ by 
  \begin{equation}
  d_i=\mathcal{R}_D(\mathcal{M}_{i-1},\;\theta_i)
  \end{equation}
  where $\theta_i$ is the camera parameter of the $i$-th view. Then $d_i$ serves as a reference for the more precise prediction of $D_i$. We get $D_i$ by 
  \begin{equation}
  D_i=\left\{\begin{array}{lc}t(S_i,\;\theta_i)&i=0;\\t^*(S_i,\;d_i,\;\theta_i)&i\geq 1,\end{array}\right. 
  \label{eq1}
  \end{equation}
  where $t(\cdot)$ is the initial Sketch2Depth translation model and $t^*(\cdot)$ is the enhanced Sketch2Depth translation model incorporating iterative depth reference.

\subsection{Geometry-aligned Feature Tensor Encoding}
\label{sec:encoding}
  \begin{figure*}[!t]
  \centering
  \includegraphics[width=0.9\linewidth]{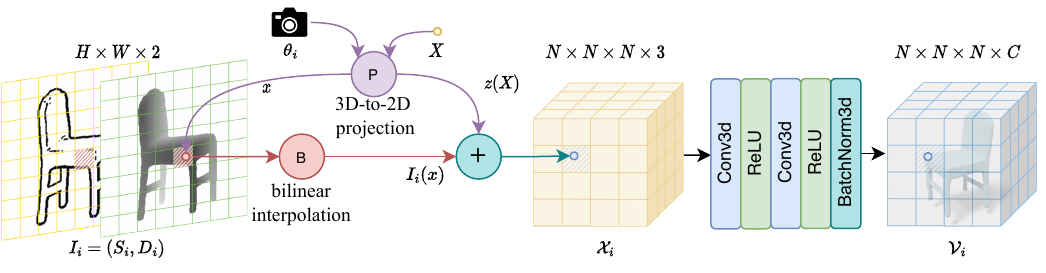}
  \caption{\small{The illustration of our geometry-aligned feature tensor encoding. We first uniformly sample $N\times N\times N$ 3D points within the 3D space. For any point $X$, let $x$ be its 2D projection location on the camera plane and $z(X)$ is the distance to the camera plane. We obtain 2D feature values $I_i(x)$ of point $X$ from $I_i$ by bilinear interpolation. Then we concatenate $z(X)$ and $I_i(x)$ to construct an initial 3D volume tensor. Then we feed $\mathcal{X}_i$ into a two-layer 3D convolutional network to learn the geometry-aligned feature tensor $\mathcal{V}_i$.}}
  \label{fig:lift}
\end{figure*}
  \noindent{\textit{\textbf{Feature Lifting.}}} Given 2D features $I_i=\{S_i, D_i\} \in \mathbb{R}^{H \times W \times 2}$ (i.e., sketches and predicted depth maps), the role of our lifting encoder is to map $I_i$ to a canonical 3D feature space $\mathbb{R}^{N\times N\times N\times C}$, where $C$ is the feature dimension. We first uniformly sample $N\times N\times N$ 3D points within the 3D space. For any point $X$, let $x$ be its 2D projection location on the camera plane, and $z(X)$ be its distance to the camera plane. We obtain 2D feature values $I_i(x)$ of point $X$ from $I_i$ by bilinear interpolation. Following PIFu~\cite{saito2019pifu}, we concatenate $I_i(x)$ and $z(X)$ to construct an initial 3D volume tensor $\mathcal{X}_i$ of size $N\times N\times N\times 3$. Then we feed $\mathcal{X}_i$ into a two-layer 3D convolutional network (as shown in Figure~\ref{fig:lift}) to learn the geometry-aligned feature tensor $\mathcal{V}_i \in \mathbb{R}^{N\times N\times N\times C}$. We use $N=64$ and $C=16$ to achieve interactive speed. 

  \noindent{\textit{\textbf{Iterative Feature Aggregation.}}} We propose an iterative feature aggregation module to support iterative shape refinement. Let $\mathcal{A}_i$ be the aggregated feature tensor of the $i$-th view. Then we have
  \begin{equation}
  \mathcal{A}_i=\left\{\begin{array}{lc}\mathcal{V}_0&i=0;\\agg(\mathcal{A}_{i-1},\;\mathcal{V}_i)&i\geq 1,\end{array}\right. 
  \label{eq2}
  \end{equation}
  where $agg(\cdot)$ is our iterative feature aggregation module. The aggregation module is composed of three stacked 3D convolutional blocks. We first concatenate the aggregated feature tensor of the $(i-1)$-th view $\mathcal{A}_{i-1}$ and the new feature tensor of the $i$-th view $\mathcal{V}_i$ and then feed it into the aggregation module to get the aggregated feature tensor of the $i$-th view $\mathcal{A}_{i}$.

\begin{figure}[!t]
  \centering
  \includegraphics[width=0.55\linewidth]{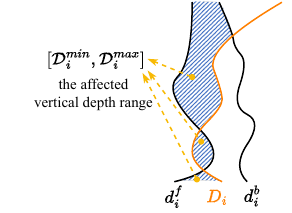}
  \caption{\small{The 2D illustration of the affected vertical depth range. The blue shaded area indicates the affected vertical depth range.}}
  \label{fig:range}
\end{figure}

  \noindent{\textit{\textbf{Local Editing with 3D Mask.}}} For shape editing, we propose a masked editing strategy to directly update the features in the editing region without affecting the features in the remaining region. Specifically, we ask the user to mark out a 2D binary mask ${M}_i^{2D}$ (for each pixel, $1$ for editing and $0$ for non-editing) and modify the sketch only within the editing region. Then, we need to determine the corresponding 3D editing mask ${M}_i^{3D}$. Just like we obtain the volume of a cylinder by multiplying its base area with its height, we obtain the 3D editing mask by multiplying the 2D editing mask and the affected vertical depth range. As described in Section~\ref{sec:prior}, we can get the predicted depth $D_i$ by our Sketch2Depth translator. Also, we can use our depth renderer $\mathcal{R}_D$ to render a front depth $d_i^f$ and a back depth $d_i^b$ from $\mathcal{M}_{i-1}$, and they satisfy the following relationship: $0 < d_i^f \leq d_i^b \leq 1$. For computational convenience, we normalize all depth values between $0$ and $1$ ($0$ means where the camera is, and $1$ means the farthest from the camera). As illustrated in Figure~\ref{fig:range}, $D_i$, $d_i^f$, and $d_i^b$ together determine the affected vertical depth range $[\mathcal{D}_i^{min}, \mathcal{D}_i^{max}]$, where $\mathcal{D}_i^{min}$ is the affected starting depth and $\mathcal{D}_i^{max}$ is the affected ending depth. Finally we get ${M}_i^{3D}$ by
  \begin{equation}
  \begin{split}
  M_i^{3D} &= M_i^{2D} * (\mathcal{D}_i^{max}-\mathcal{D}_i^{min}), \\
  \mathcal{D}_i^{min}&=min(D_i,\;d_i^f), \\
  \mathcal{D}_i^{max}&=min(max(D_i,\;d_i^f),\;d_i^b).\\
  \end{split}
  \end{equation}
  To mitigate the boundary disturbance caused by the direct feature replacement operation, we perform morphological dilation $dilate(\cdot)$ on $M_i^{3D}$ to get a dilated 3D mask $\hat{M}_i^{3D}$ with the kernel size $k$ (we use $k=5$ in our implementation):
  \begin{equation}
  \hat{M}_i^{3D} = dilate(M_i^{3D},\;k).
  \label{eq3}
  \end{equation} 
  Finally, we update the old feature tensor $\mathcal{A}_{i-1}$ to a new feature tensor $\mathcal{A}_i$ by replacing the feature within $\hat{M}_i^{3D}$:
  \begin{equation}
  \begin{split}
  \mathcal{A}_i'&=(1-\hat{M}_i^{3D})\cdot\mathcal{A}_{i-1}+\hat{M}_i^{3D}\cdot\mathcal{V}_i, \\
  \mathcal{A}_i''&=agg(\mathcal{A}_i',\;\mathcal{V}_i), \\
  \mathcal{A}_i&=(1-\hat{M}_i^{3D})\cdot\mathcal{A}_{i-1}+\hat{M}_i^{3D}\cdot\mathcal{A}_i''.
  \end{split}
  \label{eq4}
  \end{equation}
  Note that we use a sandwich-like \textit{replace-aggregate-replace} strategy to improve the robustness of masked editing. The first replacement operation replaces old features within the editing region with new features. The next aggregation operation is intended to smooth the neighborhood surrounding the editing region. No further training is needed, and we use the same $agg(\cdot)$ as Section~\ref{sec:encoding}. The following second replacement operation aims to keep the remaining region unchanged.
  
\subsection{Implicit Surface Prediction}
\label{sec:prediction}
  We represent the surface as the occupied/unoccupied space decision boundary of continuous occupancy fields and apply the Marching Cube algorithm~\cite{lorensen1987marching} to recover the target meshes. Given the aggregated volume-based feature tensors $\mathcal{A}$ of size $64\times 64\times 64\times 16$, we use a multi-scale encoder to extract multi-scale deep feature grids inspired by IF-Net~\cite{chibane20ifnet}. Applying 3D convolution and max-pooling recursively three times on $\mathcal{A}$, we get multi-scale deep feature grids $F_0$, $F_1$, $F_2$, and $F_3$. The corresponding decreasing resolutions are $64$, $32$, $16$, and $8$. The corresponding feature channel dimensions are $32$, $64$, $128$, and $128$. Given any query point $p\in\mathbb{R}^3$, we extract the learned deep features $F_0(p)$, ..., $F_3(p)$ from $F_0$, ..., $F_3$ by trilinear interpolation. To reduce the computation cost, we make the network as simple as possible by removing the neighbor displacement scheme used by IF-Net. Finally, we concatenate $F_0(p)$, ..., $F_3(p)$ and the coordinate location together and then feed them into a point-wise decoder $f(\cdot)$, parameterized by a fully connected neural network, to predict if the query point $p$ lies inside (classification as $1$) or outside (classification as $0$) the surface. Then the surface is implicitly represented as the points on the decision boundary, $\{p\in\mathbb{R}^3|f(F_0(p), ..., F_3(p),p)=\sigma\}$ with a threshold parameter $\sigma$ (we use $\sigma=0.5$). 

\section{Experiments and Discussion}
\subsection{Implementation Details}
\noindent{\textit{\textbf{Metrics.}}}
  To measure reconstruction quality quantitatively, we consider three established metrics: 
  \begin{itemize}
  \item \textbf{IoU}: volumetric intersection over union measuring how well the defined volumes match (higher is better).

  \item \textbf{CD}: Chamfer distance measuring the accuracy and completeness of the surface (lower is better). 

  \item \textbf{NC}: normal consistency measuring the accuracy and completeness of the shape normals (higher is better).
  \end{itemize}
  
\noindent{\textit{\textbf{Datasets.}}}
  We conduct all experiments and evaluations on ShapeNet chairs and airplanes~\cite{chang2015shapenet}. Airplanes are representative objects with simple structures, and they are basically spindle-shaped. While chairs are representative objects with complex structures, and their structures have wider variations and are more difficult to reconstruct from 2D sketches. In order to compute ground truth occupancies, we make the ShapeNet models watertight and simplified through TSDF fusion~\cite{Stutz2018ARXIV}. We use the same training and test splits as 3D-R2N2~\cite{choy20163d}. We generate synthetic sketches for training by rendering depth maps using Pytorch3D~\cite{ravi2020pytorch3d} and then extracting lines using the adaptive threshold algorithm in OpenCV~\cite{opencv_library}. We also use this method to automatically generate reference sketches in our interactive system. In order to enable our method to handle continuous viewpoints, we generate training sketches from $120$ different viewpoints (the combinations of $24$ azimuth angles ($0^\circ \sim 345^\circ$) and $5$ elevation angles ($-15^\circ \sim 45^\circ$) with $15^\circ$ as interval).

\noindent{\textit{\textbf{Training Settings.}}} 
  Some training settings for our experiments are given. We train our sketch-to-depth translator similarly to pix2pix~\cite{isola2017image} does but add a normal loss to improve the accuracy of depth prediction in areas near lines. We derive the normal from the predicted depth and then compute L1 loss between the derived normal and GT normal. The normal loss provides more supervision on the edge parts of the predicted depth. We first train our single-view network, referred to as "Ours-Single", for $200$ epochs. Subsequently, we freeze all parameters of "Ours-Single" and exclusively train our aggregation module for an additional $200$ epochs. We use the Adam optimizer with the learning rate $1e-4$ and batch size $4$ in all experiments. All experiments are conducted on a single NVIDIA GeForce RTX 2080 Ti gpu.

\subsection{Qualitative Comparison}
\subsubsection{Single-view Reconstruction}
\label{sec:svr}

  As the old adage goes, "\textit{Well begun is half done.}" Selecting an appropriate viewpoint for the first sketch is of great importance since it impacts the reconstruction quality of the subsequent sketches. We assessed the performance of our method in single-view reconstruction from $28$ distinct viewpoints, as shown in Table~\ref{table:view_selection}. Based on the statistics, we can draw several noteworthy conclusions:
    \begin{itemize}
    \item For the \textit{chair} category, an elevation angle of $15^{\circ}$ is recommended, and extreme azimuth angles that are multiples of $90^{\circ}$ should be avoided.
    \item For the \textit{airplane} category, high accuracies are obtained from elevation angles of $30^{\circ}$. Users should avoid the horizontal elevation angle where the airplane wings are barely visible.
    \end{itemize}
  We can find that more informative viewpoints tend to generate better reconstruction results. We encourage the users to use informative viewpoints to reduce ambiguity by showing the 3D object with minimal foreshortening on all sides.

\begin{table*}[!t]
\centering
\small
\begin{tabular}{c|c|cccccccc}
\bottomrule[1pt]
\textbf{Category} &\diagbox{\textbf{EL}}{\textbf{AZ}} &$0^{\circ}$&$30^{\circ}$&$60^{\circ}$&$90^{\circ}$&$120^{\circ}$&$150^{\circ}$&$180^{\circ}$\\ 
\hline
\multirow{4}{*}{\textbf{Chair}} 
&$0^{\circ}$   & 0.867 & 0.290 & 0.308 & 0.599 & 0.351 & 0.371 & 1.207 \\ 
&$15^{\circ}$  & 0.771 & 0.263 & \textbf{0.221} & 0.465 & 0.241 & 0.285 & 0.974 \\ 
&$30^{\circ}$  & 0.910 & 0.271 & 0.249 & 0.451 & 0.250 & 0.278 & 0.981 \\ 
&$45^{\circ}$  & 1.207 & 0.372 & 0.331 & 0.472 & 0.307 & 0.338 & 0.973 \\ 
\hline
\multirow{4}{*}{\textbf{Airplane}}
&$0^{\circ}$   & 0.979 & 0.461 & 0.719 & 0.587 & 0.454 & 0.548 & 1.121\\
&$15^{\circ}$  & 0.337 & 0.137 & 0.191 & 0.224 & 0.187 & 0.155 & 0.342 \\ 
&$30^{\circ}$  & 0.287 & \textbf{0.125} & 0.149 & 0.193 & 0.146 & 0.142 & 0.312 \\ 
&$45^{\circ}$  & 0.363 & 0.182 & 0.159 & 0.187 & 0.170 & 0.189 & 0.374 \\ 
\toprule[1pt]
\end{tabular}
\caption{\small{Chamfer distance ($\times 10^{-3}$) for single-view reconstruction from different viewpoints. AZ and EL represent the azimuth and elevation angles.}}
\label{table:view_selection}
\end{table*}

\begin{figure*}[!t]
  \centering
  \includegraphics[width=0.95\linewidth]{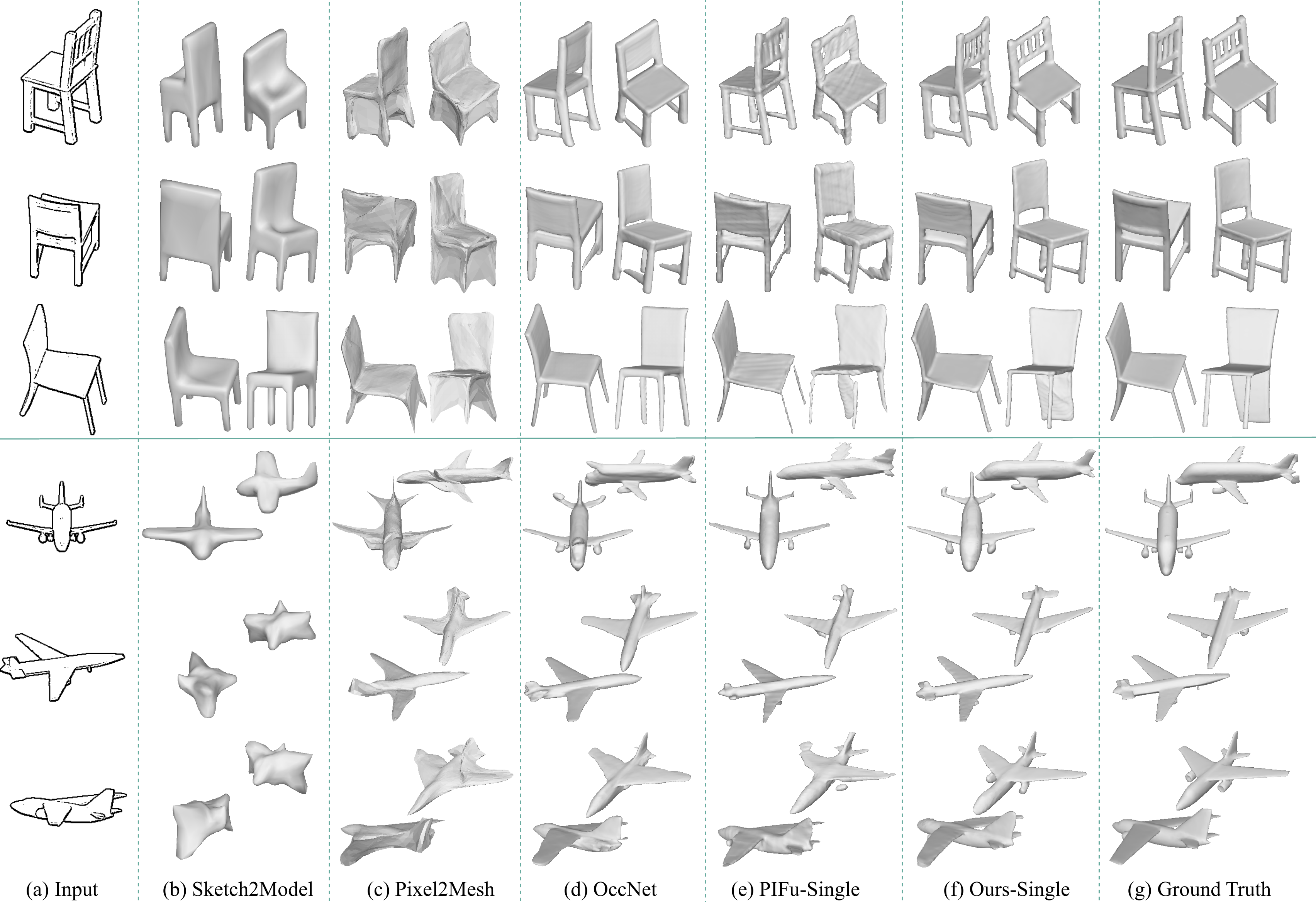}
  \caption{\small{Qualitative comparisons of our method with four previous methods for Single-view Reconstruction. "PIFu-Single" and "Ours-Single" refer to the single-view versions of "PIFu" and "Ours".}}
  \label{fig:svr}
\end{figure*}

  To evaluate the superiority of our method on single-view reconstruction, we selected four existing representative methods that use different feature encoding techniques (i.e., global or local features) and different representations of the 3D shape (i.e., explicit mesh or implicit function) for comparison: Sketch2Model~\cite{zhang2021sketch2model}, Pixel2Mesh~\cite{wang2018pixel2mesh}, OccNet~\cite{mescheder2019occupancy}, and single-view PIFu~\cite{saito2019pifu}. The experiment was conducted using the open-source code provided by these methods. Table~\ref{table:single_comparison} and Figure~\ref{fig:svr} show the results of this comparison. As seen from Table~\ref{table:single_comparison}, methods based on implicit functions (i.e., OccNet, PIFu, and ours) have superior representation abilities compared to those based on explicit mesh deformation (i.e., Sketch2Model and Pixel2Mesh). Furthermore, it shows that our method's use of geometry-aligned features results in superior performance compared to PIFu's use of pixel-aligned features. Figure~\ref{fig:svr} demonstrates that local feature-based methods (i.e., Pixel2Mesh, PIFu, and ours) produce more precise surfaces than global feature-based methods (i.e., Sketch2Model and OccNet). Our method takes advantage of both implicit representations and local geometry-aligned features, which results in superior performance compared to all existing methods in generating accurate 3D geometries. Both qualitative and quantitative results demonstrate the superiority of our method.
  
  \begin{table*}[!t]
\centering
\small
\begin{tabular}{l|l|l|c|c|c|c|c|c}
\bottomrule[1pt]
\multirow{2}{*}{\textbf{Method}} & \multicolumn{1}{c|}{\multirow{2}{*}{\textbf{Representation}}} & \multicolumn{1}{c|}{\multirow{2}{*}{\textbf{Feature Encoding}}} & \multicolumn{3}{c|}{\textbf{Chair}} & \multicolumn{3}{c}{\textbf{Airplane}} \\ 
\cline{4-9} 
& & & \textbf{IoU$\uparrow$} & \textbf{CD$\downarrow$} & \textbf{NC$\uparrow$} & \textbf{IoU$\uparrow$} & \textbf{CD$\downarrow$} & \textbf{NC$\uparrow$} \\ 
\hline
Sketch2Model~\cite{zhang2021sketch2model}& mesh deformation  & global (latent code vector) & 0.210 & 4.153 & 0.657 & 0.292 & 2.503 & 0.691 \\ 
Pixel2Mesh~\cite{wang2018pixel2mesh}     & mesh deformation  & local (pixel-aligned)       & 0.280 & 0.817 & 0.739 & 0.487 & 0.381 & 0.776 \\  \hline
OccNet~\cite{mescheder2019occupancy}     & implicit function & global (latent code vector) & 0.534 & 5.661 & 0.719 & 0.520 & 0.434 & 0.839 \\
PIFu-Single~\cite{saito2019pifu}         & implicit function & local (pixel-aligned)       & 0.582 & 0.546 & 0.838 & 0.656 & 0.433 & 0.855 \\ \hline
Ours-Single                              & implicit function & local (geometry-aligned)    & \textbf{0.590} & \textbf{0.378} & \textbf{0.853} & \textbf{0.703} & \textbf{0.330} & \textbf{0.879} \\
\toprule[1pt]
\end{tabular}
\caption{\small{Single-view quantitative evaluations of different methods on ShapeNet chairs and airplanes. The unit of CD is $10^{-3}$. "PIFu-Single" and "Ours-Single" refer to the sing-view versions of "PIFu" and "Ours".}}
\label{table:single_comparison}
\end{table*}

\subsubsection{Multi-view Reconstruction}
\label{sec:mvr}
\begin{figure*}[!t]
  \centering
  \includegraphics[width=0.99\linewidth]{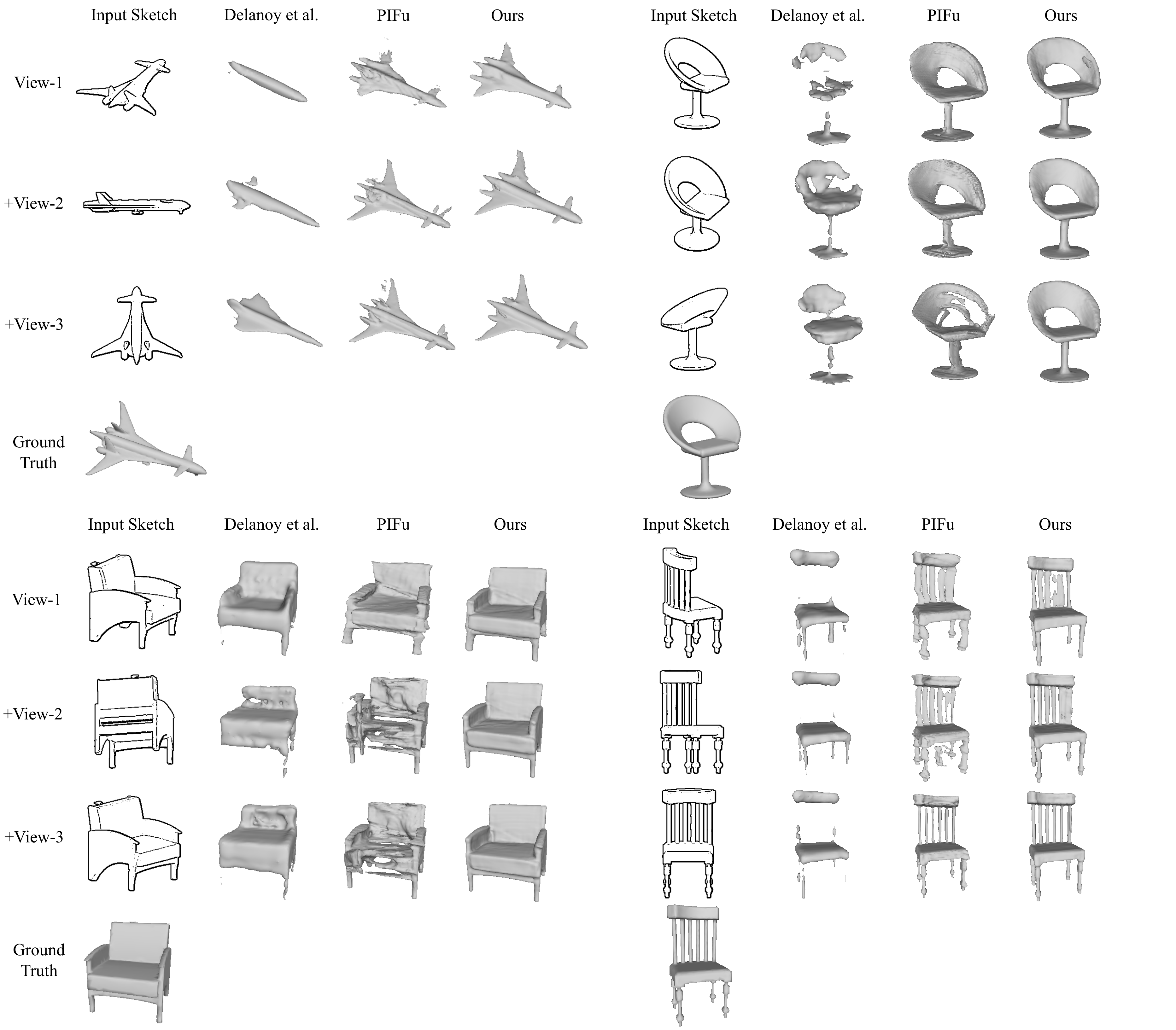}
  \caption{\small{Qualitative comparisons of our method with two previous methods for iterative Multi-view Reconstruction. The sketches are rendered from the ground truth meshes from three random viewpoints to imitate iterative sketching. \textbf{In each case, the first row displays the results obtained from View-1 sketch, while the second row shows the results from View-1 and View-2 sketches. The third row displays the results from View-1, View-2, and View-3 sketches.}}}
  \label{fig:mvr}
\end{figure*}

  To verify the performance of our method on multi-view reconstruction, we compare our approach with two methods most related to our method: Delanoy et al.~\cite{delanoy20183d} and multi-view PIFu~\cite{saito2019pifu}. We use 3 sketches from randomly selected viewpoints to evaluate each method. Based on the recommended viewpoints discussed in Section~\ref{sec:svr} and the suggestion of Delanoy et al. (imposing a 3/4 view as the first drawing to significantly reduce ambiguity), the viewpoint of the first sketch is selected from some informative viewpoints, avoiding accidental viewpoints. For the method proposed by Delanoy et al., we re-implement their method and perform 5 iterations for refinement. 
  For multi-view PIFu, we adopt their open-source code (\href{https://github.com/shunsukesaito/PIFu}{https://github.com/shunsukesaito/PIFu}). Table~\ref{table:multiview_comparison} presents the quantitative comparisons of the three methods mentioned above.  We also provide qualitative comparisons in Figure~\ref{fig:mvr}. Both the quantitative and qualitative results indicate that our method achieves the highest performance in geometry inference and produces reconstructions that are closer to the ground truth.
  \begin{table*}[!t]
\centering
\small
\begin{tabular}{l|l|c|c|c|c|c|c|c|c|c}
\bottomrule[1pt]
\multicolumn{1}{c|}{\multirow{2}{*}{\textbf{Category}}} & \multirow{2}{*}{\textbf{Method}} & \multicolumn{3}{c|}{\textbf{1st View}} & \multicolumn{3}{c|}{\textbf{2nd View}} & \multicolumn{3}{c}{\textbf{3rd View}} \\ \cline{3-11}
& & \textbf{IoU$\uparrow$} & \textbf{CD$\downarrow$} & \textbf{NC$\uparrow$} & \textbf{IoU$\uparrow$} & \textbf{CD$\downarrow$} & \textbf{NC$\uparrow$} & \textbf{IoU$\uparrow$} & \textbf{CD$\downarrow$} & \textbf{NC$\uparrow$}\\  
\hline
\multirow{3}{*}{\textbf{Chair}} 
& Delanoy et al.~\cite{delanoy20183d} & 0.413 & 3.232 & 0.777 & 0.387 & 4.553 & 0.761 & 0.385 & 3.914 & 0.762 \\
& PIFu~\cite{saito2019pifu}           & 0.570 & 0.375 & 0.814 & 0.606 & 1.585 & 0.805 & 0.674 & 0.553 & 0.846 \\
& Ours                                & \textbf{0.631} & \textbf{0.230} & \textbf{0.876} & \textbf{0.705} & \textbf{0.160} & \textbf{0.904} & \textbf{0.741} & \textbf{0.124} & \textbf{0.918} \\

\hline
\multirow{3}{*}{\textbf{Airplane}} 
& Delanoy et al.~\cite{delanoy20183d} & 0.562 & 2.090 & 0.814 & 0.556 & 1.975 & 0.818 & 0.562 & 2.384 & 0.817 \\
& PIFu~\cite{saito2019pifu}           & 0.637 & 0.236 & 0.849 & 0.727 & 0.275 & 0.865 & 0.778 & 0.160 & 0.890 \\
& Ours                                & \textbf{0.730} & \textbf{0.131} & \textbf{0.896} & \textbf{0.797} & \textbf{0.083} & \textbf{0.916} & \textbf{0.821} & \textbf{0.077} & \textbf{0.924} \\ 

\toprule[1pt]
\end{tabular}
\caption{\small{Multi-view quantitative evaluations of different methods on ShapeNet chairs and airplanes. The unit of CD is $10^{-3}$.}}
\label{table:multiview_comparison}
\end{table*}

\subsection{Ablation Study}

\subsubsection{Effectiveness of Depth Prior}
  \begin{table*}[!t]
\centering
\small
\begin{tabular}{l|l|c|c|c|c|c|c|c|c|c}
\bottomrule[1pt]
\multicolumn{1}{c|}{\multirow{2}{*}{\textbf{Category}}} & \multirow{2}{*}{\textbf{Mehtod}} & \multicolumn{3}{c|}{\textbf{1st View}} & \multicolumn{3}{c|}{\textbf{2nd View}} & \multicolumn{3}{c}{\textbf{3rd View}} \\ \cline{3-11}
& & \textbf{IoU$\uparrow$} & \textbf{CD$\downarrow$} & \textbf{NC$\uparrow$} & \textbf{IoU$\uparrow$} & \textbf{CD$\downarrow$} & \textbf{NC$\uparrow$} & \textbf{IoU$\uparrow$} & \textbf{CD$\downarrow$} & \textbf{NC$\uparrow$}\\
\hline
\multirow{2}{*}{\textbf{Chair}}
& Ours-no-DP   & 0.521 & 0.619 & 0.798 & 0.655 & 0.247 & 0.875 & 0.695 & 0.180 & 0.894 \\ 
& Ours        & \textbf{0.631} & \textbf{0.230} & \textbf{0.876} & \textbf{0.705} & \textbf{0.160} & \textbf{0.904} & \textbf{0.741} & \textbf{0.124} & \textbf{0.918} \\
\hline
\multirow{2}{*}{\textbf{Airplane}} 
& Ours-no-DP   & 0.666 & 0.292 & 0.862 & 0.764 & 0.184 & 0.903 & 0.787 & 0.070 & 0.913 \\ 
& Ours        & \textbf{0.730} & \textbf{0.131} & \textbf{0.896} & \textbf{0.797} & \textbf{0.083} & \textbf{0.916} & \textbf{0.821} & \textbf{0.077} & \textbf{0.924} \\ 
\toprule[1pt]
\end{tabular}
\caption{\small{Ablation studies about the effectiveness of depth prior. "Ours-no-DP" refers to a variant of "Ours" in which the depth prediction branch has been removed. The unit of CD is $10^{-3}$.}}
\label{table:ablation_dp}
\end{table*}
  To show the effectiveness of our view-aware depth prior, we remove the depth prediction branch as a vanilla version called "Ours-no-DP". Table~\ref{table:ablation_dp} shows that depth prior does bring some nice-to-have improvements by offering richer features. On the other hand, our method works well even for only sparse sketch information.
 
\subsubsection{Effectiveness of Iterative Feature Aggregation}
  \begin{table*}[!t]
\centering
\small
\begin{tabular}{l|l|c|c|c|c|c|c|c|c|c}
\bottomrule[1pt]
\multicolumn{1}{c|}{\multirow{2}{*}{\textbf{Category}}} & \multirow{2}{*}{\textbf{Mehtod}} & \multicolumn{3}{c|}{\textbf{1st View}} & \multicolumn{3}{c|}{\textbf{2nd View}} & \multicolumn{3}{c}{\textbf{3rd View}} \\ \cline{3-11}
& & \textbf{IoU$\uparrow$} & \textbf{CD$\downarrow$} & \textbf{NC$\uparrow$} & \textbf{IoU$\uparrow$} & \textbf{CD$\downarrow$} & \textbf{NC$\uparrow$} & \textbf{IoU$\uparrow$} & \textbf{CD$\downarrow$} & \textbf{NC$\uparrow$}\\ 
\hline
\multirow{3}{*}{\textbf{Chair}} 
& Ours-AVG        & 0.631 & 0.230 & 0.876 & 0.697 & 0.164 & 0.904 & 0.732 & 0.132 & \textbf{0.918} \\
& Ours-Updater    & 0.631 & 0.230 & 0.876 & 0.699 & \textbf{0.158} & \textbf{0.905} & 0.726 & \textbf{0.122} & \textbf{0.918} \\
& Ours            & 0.631 & 0.230 & 0.876 & \textbf{0.705} & 0.160 & 0.904 & \textbf{0.741} & 0.124 & \textbf{0.918} \\
\hline
\multirow{3}{*}{\textbf{Airplane}} 
& Ours-AVG        & 0.730 & 0.131 & 0.896 & 0.790 & 0.096 & 0.916 & 0.815 & 0.083 & \textbf{0.924} \\
& Ours-Updater    & 0.730 & 0.131 & 0.896 & 0.788 & 0.086 & \textbf{0.917} & 0.797 & 0.079 & 0.922 \\
& Ours            & 0.730 & 0.131 & 0.896 & \textbf{0.797} & \textbf{0.083} & 0.916 & \textbf{0.821} & \textbf{0.077} & \textbf{0.924} \\
\toprule[1pt]
\end{tabular}
\caption{\small{Ablation studies about the effectiveness of iterative feature aggregation. "Ours-AVG" and "Ours-Updater" refer to two variants of "Ours" in which the iterative feature aggregation module has been replaced by average pooling operation and iterative updater (proposed by Delanoy et,al.~\cite{delanoy20183d}) respectively. The unit of CD is $10^{-3}$.}}
\label{table:ablation_fusion} 
\end{table*}
  In order to show the effectiveness of our iterative feature aggregation module, we compare it with average pooling used by~\cite{saito2019pifu} and the iterative updater proposed by Delanoy et al.~\cite{delanoy20183d}. We replace our iterative feature aggregation module with average pooling and the iterative updater~\cite{delanoy20183d}. We call the two variants "Ours-AVG" and "Ours-Updater". For a fair comparison, "Ours-AVG", "Ours-Updater" and "Ours" share the same lifting encoder parameters. It can be observed from Table~\ref{table:ablation_fusion} that taking the three metrics into consideration, "Ours" achieves a greater improvement than two variants as the number of viewpoints increases. This means our iterative feature aggregation module has better support for iterative modeling. 
  
\subsubsection{Different Initial Sketches}
\begin{figure}[!t]
  \centering
  \includegraphics[width=0.9\linewidth]{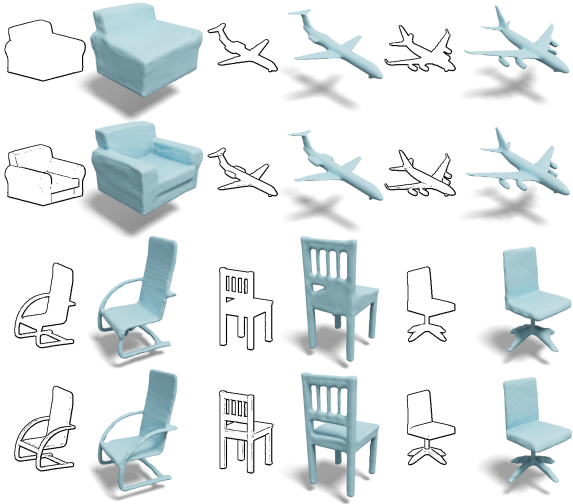}
  \caption{\small{Some examples when provided with "contour-only" and "suggestive-contour" sketches.}}
  \label{fig:contour}
\end{figure}

\begin{figure}[!t]
  \centering
  \includegraphics[width=0.99\linewidth]{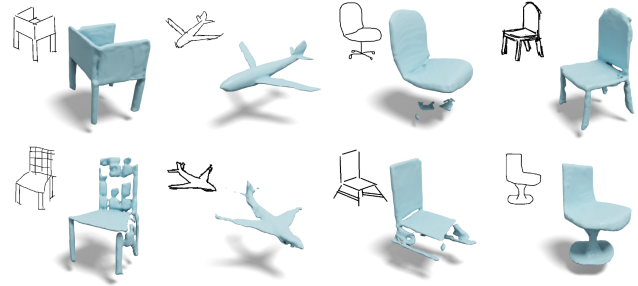}
  \caption{\small{Some examples of freehand-sketch-conditioned shape generation of our method.}}
  \label{fig:freehand}
\end{figure}

  As an interactive system, the robustness to different types of initial sketches needs to be considered. In Figure~\ref{fig:contour}, we compare the results when provided with "contour only" and "suggestive-contour" sketches. Our method can generate reasonable results even with "contour-only" sketches lacking internal structures. Nevertheless, we encourage users to draw more details. In Figure~\ref{fig:freehand}, we provide some freehand-sketch-conditioned results. Because we train our networks with synthesis sketches, which have a significant domain gap with freehand sketches, our method may not perform well with too abstract sketches. However, our method still tends to generate faithful results that reflect users' real intentions. While the initial shape reconstruction may be poor when provided with a poorly drawn initial sketch, our interactive system allows users to refine or edit the reconstructed surface iteratively. 

\subsection{User Study}
  We have conducted two informal user studies, i.e., the Usability Study and the Perceptive Study, to evaluate the usability and effectiveness of our system and the quality of results generated by our method.

\subsubsection{Usability Study}
\begin{figure*}[!t]
  \centering
  \includegraphics[width=0.8\linewidth]{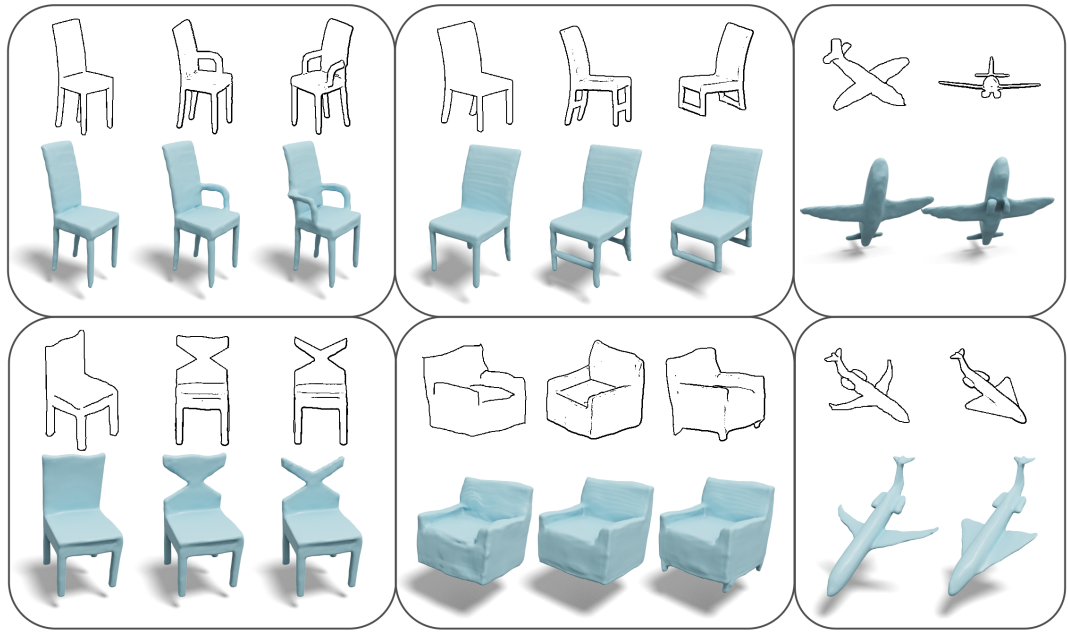}
  \caption{\small{The gallery of our results. All models are created by amateur users who are trained to use our system with a tutorial. Thanks to the easy-to-use modeling design, most users can complete each model design with 2-3 views.}}
  \label{fig:gallery}
\end{figure*}

  The main purpose of our work is to design a user-friendly 3D modeling system specifically tailored for amateur users without 3D modeling experience. To verify the usability of our system, we recruited 8 participants (P1-P8), aged 18 to 28, from various departments at a local university. Based on the pre-study survey, none of these participants had experience in 3D modeling. Four of them (P2, P3, P6, P8) had professional 2D drawing experience, and the rest had limited drawing experience. Before the usability study, each participant was provided with an 8-minute video demonstrating the basic operations of our system. Subsequently, each user was given 15 minutes to become familiar with our system. After the tutorial session, each participant was tasked with freely creating at least two models (a chair and an airplane) without any constraint on diversity, quality or drawing time. Figure~\ref{fig:gallery} shows representative results created by these participants with our system. It can be seen from this figure that our system supports amateur users to create objects with diversified shapes.

\begin{figure}[!t]
  \centering
  \includegraphics[width=0.99\linewidth]{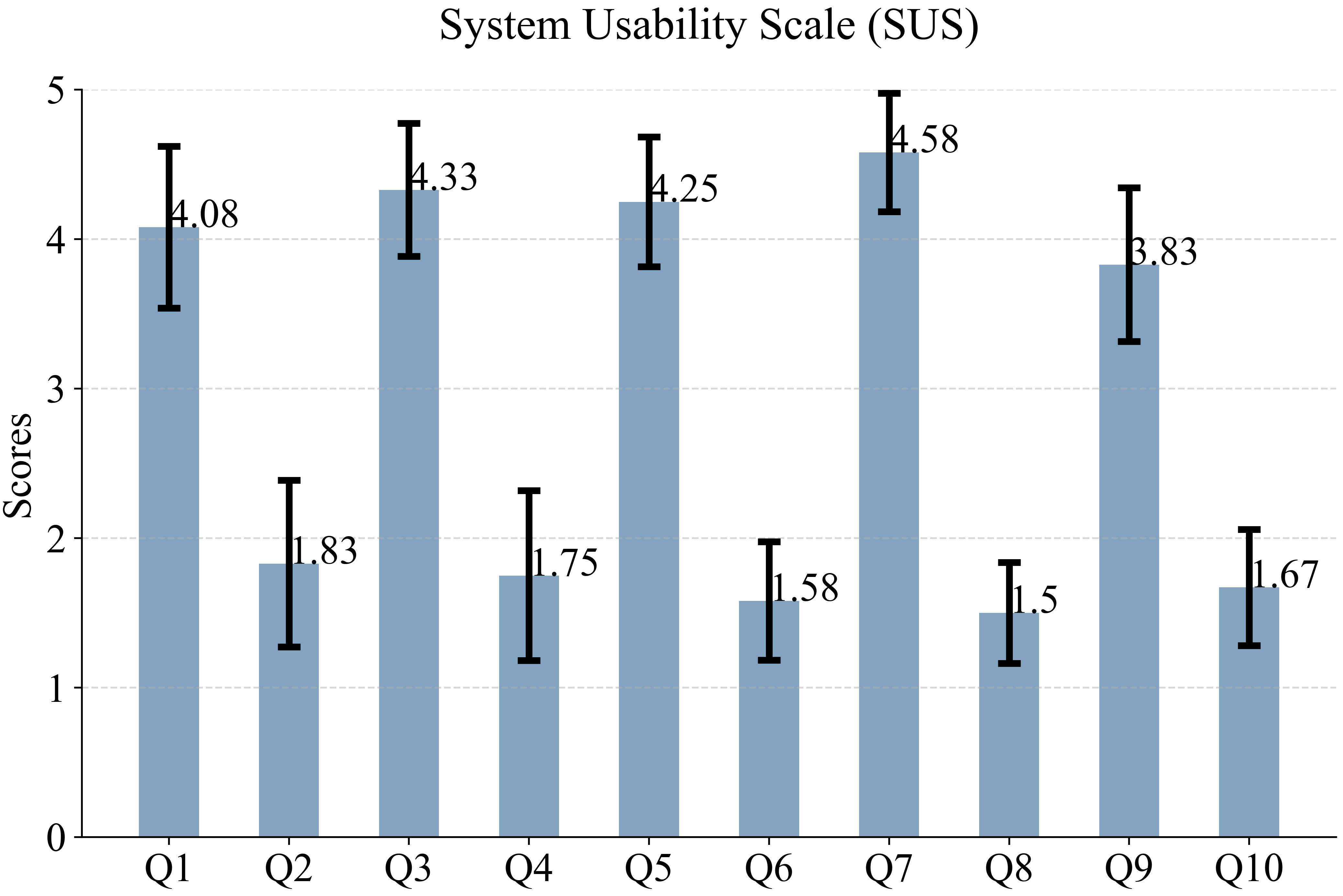}
  \caption{\small{Mean scores of SUS in a 5-point scale. For the questions with the odd numbers, the higher the scores are the better; for the rest of the questions, the lower the scores are the better.}}
  \label{fig:sus}
\end{figure}
\begin{figure}[htbp]
  \centering
  \includegraphics[width=0.9\linewidth]{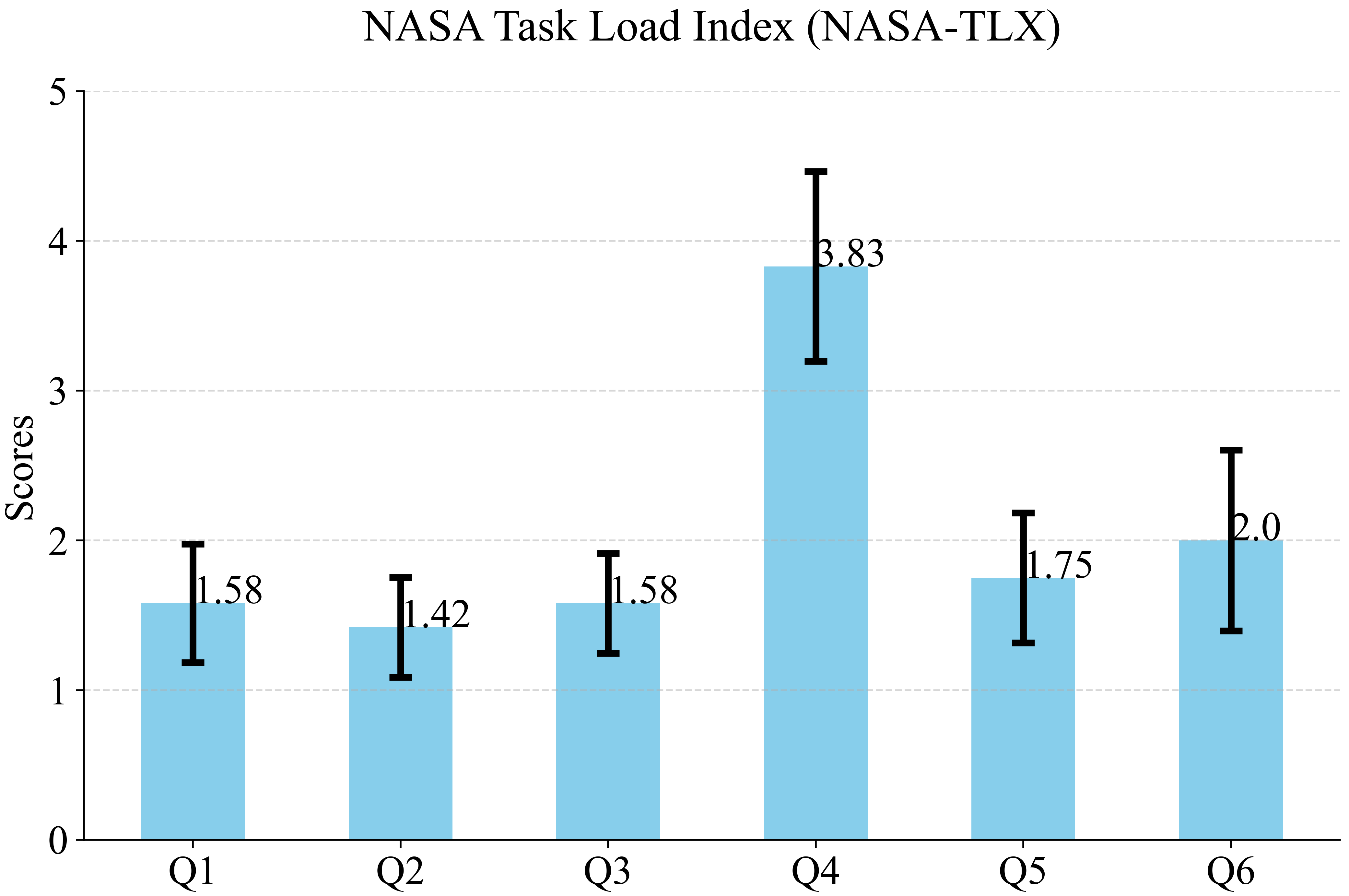}
  \caption{\small{Mean scores of NASA-TLX in a 5-point scale. Q1: Mental Demand, Q2: Physical Demand, Q3: Temporal Demand, Q4: Performance, Q5: Effort, Q6: Frustration.}}
  \label{fig:nasa}
\end{figure}

  At the end of the modeling session, each participant was required to fill in a System Usability Scale (SUS) questionnaire and a NASA Task Load Index (NASA-TLX) questionnaire to evaluate the usability and workload of our system. Figure~\ref{fig:sus} illustrates the average score for each question in the SUS questionnaire. For the questions with the odd numbers, the higher the SUS scores, the better; for the rest of the questions, the lower the SUS scores, the better. The high scores of Q3, Q5, Q7, as well as the low scores of Q2, Q4, Q6, Q8, and Q10, suggest that our system effectively supports amateur users in easily and intuitively creating 3D objects, indicating good user-friendliness and usability of our system. The high scores of Q1 and Q9 show that most participants were satisfied with the models generated by our system. Figure~\ref{fig:nasa} illustrates the mean score for each question in the NASA-FLX questionnaire. The results for NASA-TLX are also positive, as indicated by the extremely low levels of mental demand, physical demand, temporal demand, effort, and frustration, along with a relatively high-performance score. It suggests that our system enables users to effortlessly customize desired 3D objects.

\subsubsection{Perceptive Study}
\begin{figure}[!t]
  \centering
  \includegraphics[width=0.8\linewidth]{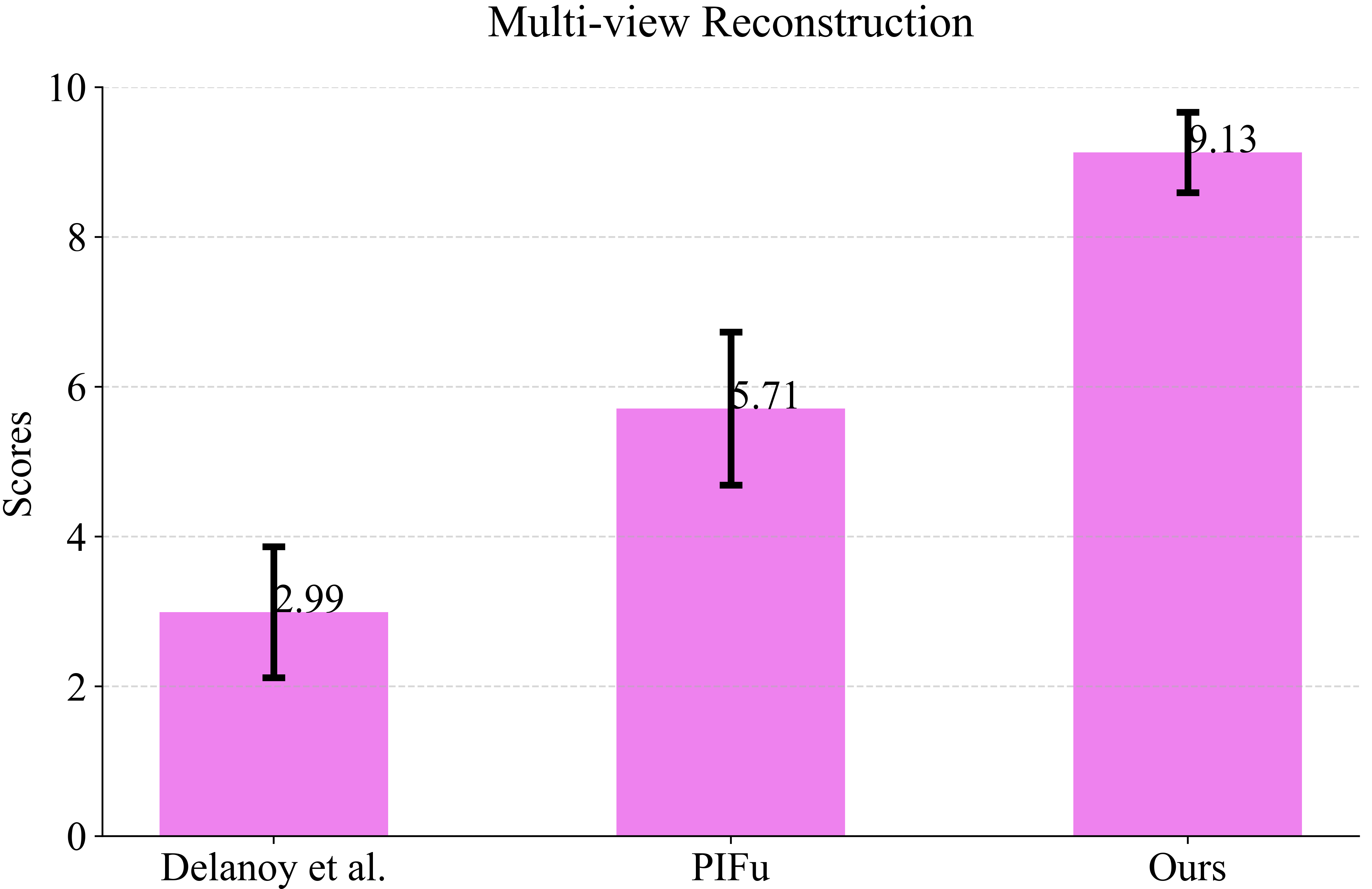}
  \caption{\small{Mean scores of Perceptive Study on Multi-view Reconstruction in a 10-point scale. The higher the scores are the better. }}
  \label{fig:mvr_chart}
\end{figure}

\begin{figure}[!t]
  \centering
  \includegraphics[width=0.9\linewidth]{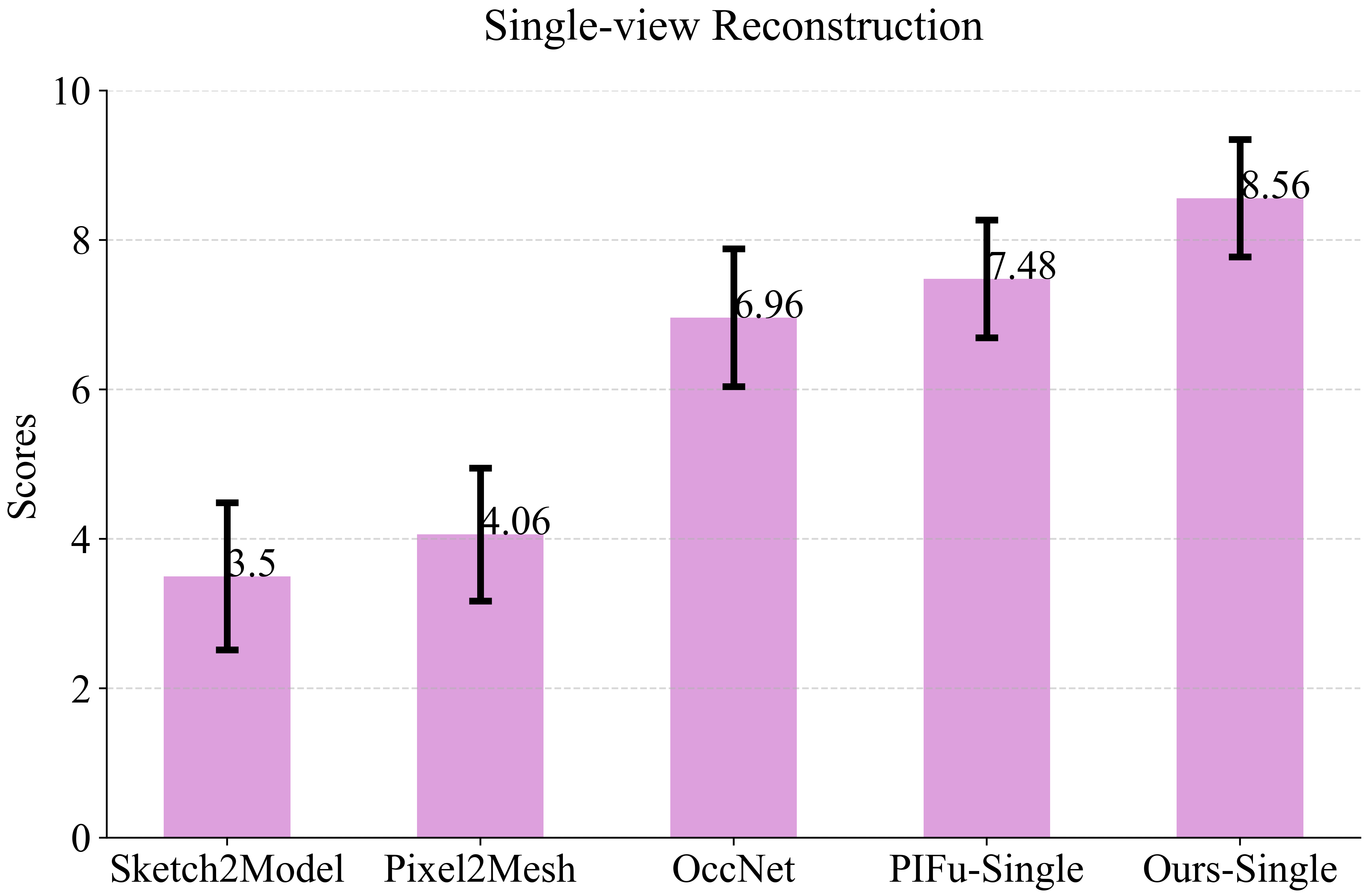}
  \caption{\small{Mean scores of Perceptive Study on Single-view Reconstruction in a 10-point scale. The higher the scores are the better.}}
  \label{fig:svr_chart}
\end{figure}
  To further evaluate the effectiveness and superiority of our method, we conducted a perceptive/subjective user study. We invited 40 individuals to participate in this subjective evaluation through an online questionnaire. Most of them had no experience in 3D modeling, and none had taken part in the usability study. For multi-view reconstruction, we randomly selected a set of results from the test set and the usability study, which included 8 groups of multi-view sketches (4 selected from the test set and 4 drawn by users, with each group containing 2-3 views), along with the corresponding models generated by different methods in Section~\ref{sec:mvr}. For single-view reconstruction, we selected 12 sketches from the test set and the corresponding results generated by different methods discussed in Section~\ref{sec:svr}. Each case in the questionnaire showed the input single-view/multi-view sketch/sketches and the models generated by different methods, placed side by side in random order. All subjects were also asked to evaluate the faithfulness of each model (i.e., the degree of fitness to input sketch/sketches) on a ten-point Likert scale (1 = lowest fitness to 10 = highest fitness). Figure~\ref{fig:mvr_chart} and Figure~\ref{fig:svr_chart} show the results for this study. As observed from these two figures, the models generated by our method achieved significantly higher scores compared to all existing methods in both multi-view reconstruction and single-view reconstruction.


\section{Application}

\subsection{Sketch-guided Point Cloud Completion}
\begin{figure}[!htb]
  \centering
  \includegraphics[width=0.99\linewidth]{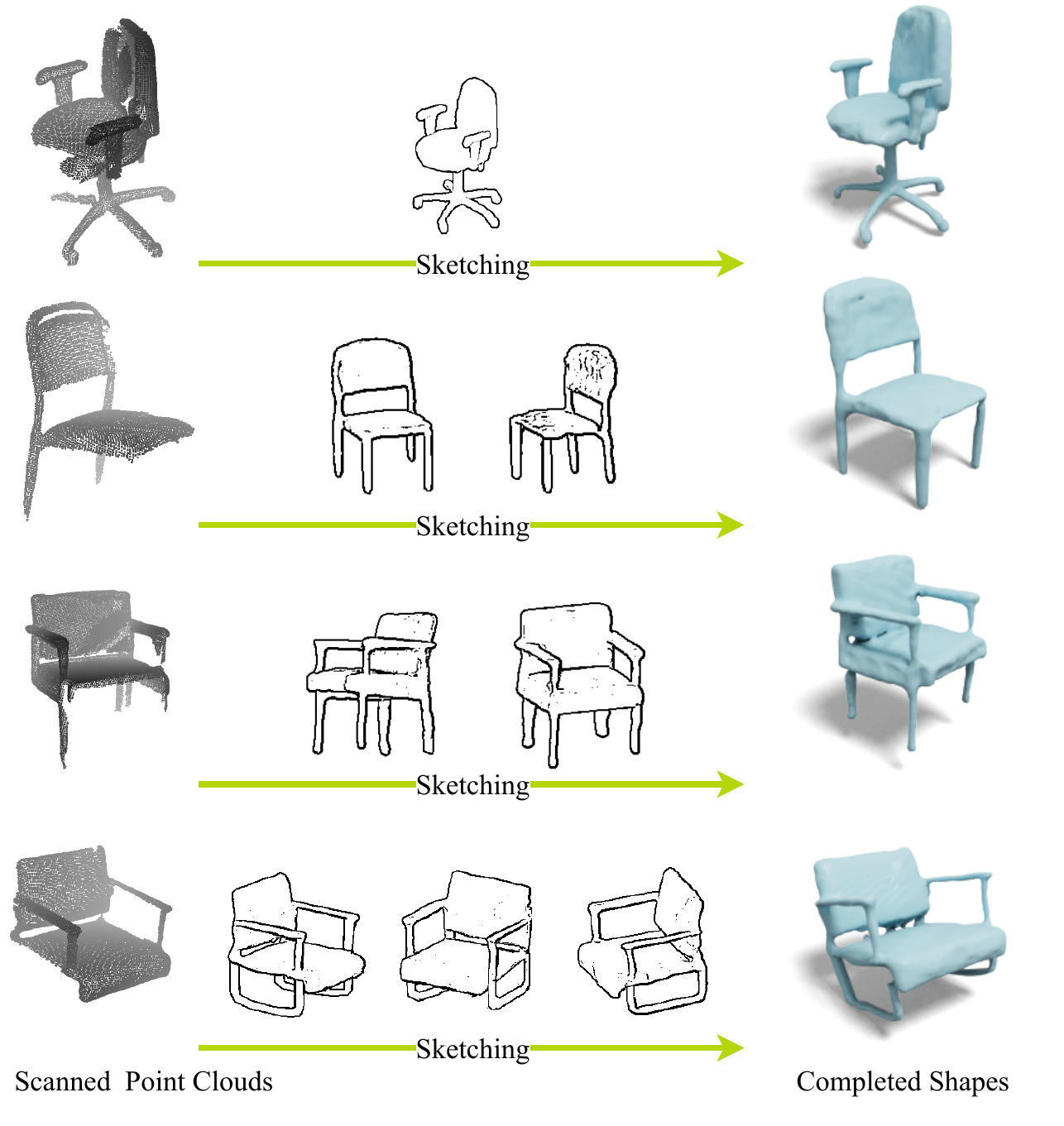}
  \caption{\small{Application on sketch-guided point cloud completion.}}
  \label{fig:app1}
\end{figure}
  Due to our geometry-aligned feature encoding, our method is perfectly compatible with the 3D shape completion task because the volume-based feature tensors are aligned with the real-world scanned point clouds. We extend our method to sketch-guided point completion by adding a voxelized point cloud as an additional feature channel (the dimension of $\mathcal{X}_i$ is $N\times N \times N \times 4$). Figure~\ref{fig:app1} shows some completed examples from scanned point clouds which are brought from ScanObjectNN~\cite{uy2019revisiting}.

\subsection{Sketch-based Cartoon Animal Head Modeling}
\begin{figure*}[!htb]
  \centering
  \includegraphics[width=0.8\linewidth]{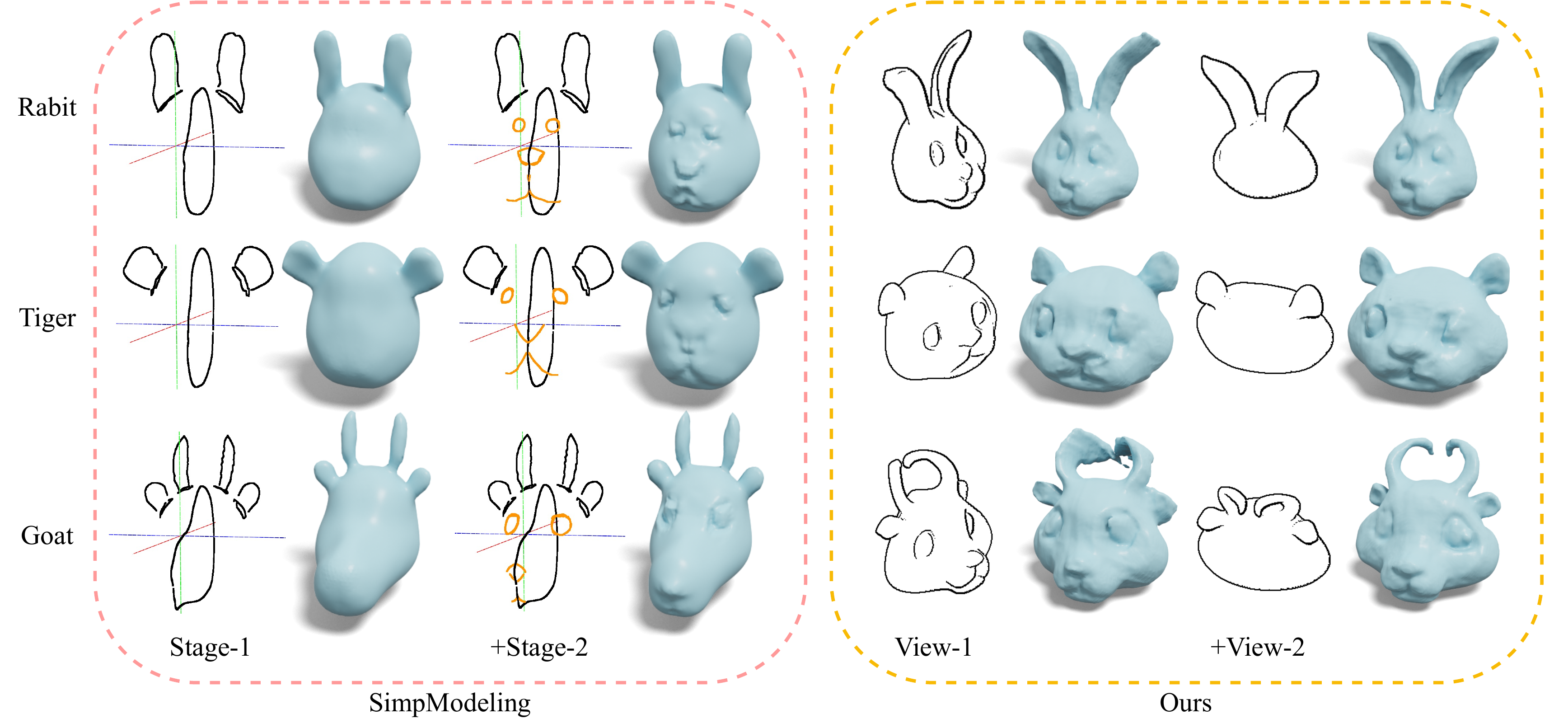}
  \caption{\small{Comparison of SimplModeling and our method for sketch-based cartoon animal head modeling.}}
  \label{fig:app2}
\end{figure*}
  We also extend our method to animal-like 3D character head modeling with iterative sketching. The sampled examples are brought from AnimalHead dataset proposed by SimpModeling~\cite{luo2021simpmodeling}. As shown in Figure~\ref{fig:app2}, we use SimpModeling and our method to model a "rabit" head, a "tiger" head and a "goat" head. Compared to SimpModeling which separates coarse shape design and geometric detail specification into two stages and respectively provide different sketching means, our system allows for intuitive 3D modeling from 2D sketches, without complex 3D interactions.

\section{Conclusion}
To sum up, we have introduced a novel iterative approach for generating 3D shapes from 2D sketches by utilizing geometry-aligned deep implicit functions. Furthermore, we have developed a unified interactive system that enables sketch-based shape generation and editing. Our extensive experiments and user study demonstrate that our proposed solution has the potential to advance the field of 3D modeling and enhance the user experience for designers, artists, and even beginners.

The limitations of our method should be acknowledged. Firstly, the robustness to imprecise sketches still need to be improved, as shown in Figure~\ref{fig:freehand}. Furthermore, the surfaces generated by our method may exhibit some noise. To enhance the quality of the reconstruction, additional post-processing techniques such as those described in~\cite{vollmer1999improved, desbrun1999implicit} can be applied. Finally, in our method, the 3D editing mask is achieved through approximate mathematical computation, and a more elegant strategy, such as using deep networks to predict the 3D editing mask, could be explored to improve this aspect of the method. In the future, we plan to improve the reliability of our system by incorporating datasets of freehand sketches. Furthermore, we may explore the possibility of texturing the generated shape using text-to-image models such as~\cite{chen2023text2tex, richardson2023texture}.

\bibliographystyle{eg-alpha-doi} 
\bibliography{my_reference}       



\end{document}